\documentclass[prb,twocolumn,aps,superscriptaddress,showpacs,longbibliography]{revtex4-2} %
\usepackage[english]{babel}
\usepackage[utf8]{inputenc}
\usepackage[colorinlistoftodos, color=green!40, prependcaption]{todonotes}
\usepackage{xcolor}
\usepackage{amsmath,amssymb} 
\usepackage{siunitx}
\usepackage{mathtools}
\usepackage{physics}
\usepackage{graphicx}
\usepackage{adjustbox}
\usepackage{placeins}
\usepackage[T1]{fontenc}
\usepackage{lipsum}
\usepackage{csquotes}
\usepackage[normalem]{ulem}

\usepackage[pdftex, pdftitle={Article}, pdfauthor={Author}]{hyperref} 
\begin{document}
\title{Probing bulk superconductivity in centrosymmetric Te-doped PtBi$_2$ single crystals}

\author{Kilian Srowik}
\thanks{These authors contributed equally to this work.}
\thanks{Current adress: Max Planck Institute for Chemical Physics of Solids,
Nöthnitzer Str. 40, 01187 Dresden, Germany}
\email{kilian.srowik@cpfs.mpg.de}
\affiliation{Leibniz Institute for Solid State and Materials Research, IFW Dresden, 01069 Dresden, Germany}

\author{Pablo Pedrazzini}
\thanks{These authors contributed equally to this work.}
\affiliation{Leibniz Institute for Solid State and Materials Research, IFW Dresden, 01069 Dresden, Germany}
\affiliation{Centro Atómico Bariloche, CNEA-CONICET, 8400 S.C. de Bariloche, Argentina}

\author{Soumen Ash}
\thanks{Current adress: Max Planck Institute for Chemical Physics of Solids,
Nöthnitzer Str. 40, 01187 Dresden, Germany}
\affiliation{Leibniz Institute for Solid State and Materials Research, IFW Dresden, 01069 Dresden, Germany}

\author{Oksana Kvitnitskaya}
\affiliation{Leibniz Institute for Solid State and Materials Research, IFW Dresden, 01069 Dresden, Germany}
\affiliation{B.~Verkin Institute for Low Temperature Physics and Engineering of the National Academy of Sciences of Ukraine, Kharkiv 61103, Ukraine}

\author{Andrii Kuibarov}
\affiliation{Leibniz Institute for Solid State and Materials Research, IFW Dresden, 01069 Dresden, Germany}

\author{Susmita Changdar}
\affiliation{Leibniz Institute for Solid State and Materials Research, IFW Dresden, 01069 Dresden, Germany}

\author{Oleksandr Suvorov}
\affiliation{Leibniz Institute for Solid State and Materials Research, IFW Dresden, 01069 Dresden, Germany}

\author{Volodymyr Bezguba}
\affiliation{Kyiv Academic University,
36 Academician Vernadsky Blvd., UA-03142 Kyiv, Ukraine}

\author{Alexander Kordyuk}
\affiliation{Kyiv Academic University,
36 Academician Vernadsky Blvd., UA-03142 Kyiv, Ukraine}

\author{Rafał Kurleto}
\affiliation{SOLARIS National Synchrotron Radiation Centre, ul. Czerwone Maki 98, 30-392 Kraków, Poland}

\author{Dawid Wutke}
\affiliation{SOLARIS National Synchrotron Radiation Centre, ul. Czerwone Maki 98, 30-392 Kraków, Poland}

\author{Reza Firouzmandi}
 \affiliation{Leibniz Institute for Solid State and Materials Research, IFW Dresden, 01069 Dresden, Germany}

\author{Robert Kluge}
\affiliation{Leibniz Institute for Solid State and Materials Research, IFW Dresden, 01069 Dresden, Germany}

\author{Swarnamayee Mishra}
 \affiliation{Institute of Solid State and Materials Physics, TU Dresden, D-01062 Dresden, Germany}  

\author{Alexander Mistonov}
\affiliation{Institute of Solid State and Materials Physics, TU Dresden, D-01062 Dresden, Germany} 

\author{Saicharan Aswartham}
\thanks{Current address: International Research Centre MagTop, Institute of Physics, Polish Academy of Sciences, al. Lotników 32/46, 02-668 Warsaw, Poland}
\affiliation{Leibniz Institute for Solid State and Materials Research, IFW Dresden, 01069 Dresden, Germany}
  
\author{Jochen Geck}
\affiliation{Institute of Solid State and Materials Physics, TU Dresden, D-01062 Dresden, Germany}
\affiliation{Institute of Solid State and Materials Physics and Würzburg-Dresden Cluster of Excellence ctd.qmat, Technische Universität Dresden, 01062 Dresden, Germany}

\author{Sergey Borisenko}
\affiliation{Leibniz Institute for Solid State and Materials Research, IFW Dresden, 01069 Dresden, Germany}

\author{Laura T. Corredor}
 \email{l.t.corredor.bohorquez@ifw-dresden.de}
\thanks{Current address: Faculty of Physics, Technical University of Dortmund, Otto-Hahn-Str. 4, D-44227 Dortmund, Germany}
 \affiliation{Leibniz Institute for Solid State and Materials Research, IFW Dresden, 01069 Dresden, Germany}

\author{Bernd Büchner}
\affiliation{Leibniz Institute for Solid State and Materials Research, IFW Dresden, 01069 Dresden, Germany}
\affiliation{Institute of Solid State and Materials Physics and Würzburg-Dresden Cluster of Excellence ctd.qmat, Technische Universität Dresden, 01062 Dresden, Germany}

\date{\today} 

\begin{abstract}

The Weyl semimetal $\gamma$-PtBi$_2$ has been shown to be one of the most promising novel materials, recently proposed as a topological \textit{i}-wave superconductor. A crucial requirement for observing this physics is the absence of inversion symmetry in its trigonal $P31m$ crystal structure. Centrosymmetry has been reported to be readily restored in the $P\overline{3}m1$ structure upon electron doping, partially substituting Bi with as little as 2\,\% Te. In this work, we synthesized Te-doped PtBi$_{2-x}$Te$_{x}$ samples and thoroughly investigated the bulk superconductivity of selected single crystals with nominal composition PtBi$_{1.96}$Te$_{0.04}$, exhibiting the highest superconducting volume fraction of $\sim100\,$\%. Single crystal XRD measurements confirm the centrosymmetric $P\overline{3}m1$ structure in our samples, whereas bulk superconductivity with a critical temperature of $T_\mathrm{c} \approx 2.4\,$K was observed in magnetization and specific heat measurements. The upper critical fields for two different orientations were determined as $H^{\parallel}_{\mathrm{c2}}(0)\approx 6.3\,$kOe for in-plane and $H^{\perp}_{\mathrm{c2}}(0)\approx 4.7\,$kOe for out-of-plane magnetic fields. Robust superconductivity was also found in resistance and point contact measurements, where the latter showed a slight enhancement of the critical temperature up to $T_\mathrm{c} \sim 3\,$K. Finally, ARPES measurements corroborate the centrosymmetric structure of our samples, showing a single surface termination and the absence of Fermi arcs.

\end{abstract}

\keywords{PtBi2, superconductivity, structural transition}
\maketitle

\section{Introduction} 
\label{sec:introduction}

Trigonal platinum bismuthide, $\gamma$-PtBi$_2$ (non-centrosymmetric space group $P31m$), has recently attracted attention as a topological system displaying remarkable properties in magnetotransport \cite{Xu2016,Gao2018,Xing2020,Wu2020,Veyrat2023} and spectroscopic measurements \cite{Thirupathaiah2018,Feng2019,Schimmel2023,Kuibarov2024, Changdar2025}. Detailed band structure calculations \cite{Veyrat2023} confirm its semimetallic character and identify the presence of Weyl nodes located only $48\,$meV above the Fermi energy. This prediction is supported by recent Angle-resolved Photoemission Spectroscopy (ARPES) measurements \cite{Kuibarov2024}, which detect surface Fermi arcs connecting Weyl nodes of opposite chirality, thereby confirming $\gamma$-PtBi$_2$ as a type I Weyl semimetal.

Furthermore, superconductivity (SC) with critical temperatures $T_\mathrm{c}\sim 0.15-1.1\,$K has been reported from electrical resistivity measurements, $\rho(T,H)$, in $\gamma$-PtBi$_2$ crystals \cite{Alekseevski1953,Shipunov2020,Zabala2024}. This SC persists even in thin exfoliated flakes, exhibiting clear 2D-SC characteristics for thicknesses up to $60\,$nm \cite{Veyrat2023}. In contrast, the superconducting state in bulk samples seems to be inhomogeneous \cite{Veyrat2022,Zabala2024}, characterized by broad transitions and very low critical currents. So far, there is no published evidence of bulk SC in $\gamma$-PtBi$_2$ as detected from thermodynamic probes. Notably, surface-sensitive spectroscopic measurements \cite{Bashlakov2022,Schimmel2023,Kuibarov2024,Besproswanny2025} reveal enhancements in both the $T_\mathrm{c}$ and the critical magnetic field $H_{c2}$.
The most remarkable results come from scanning tunneling microscopy (STM) measurements, which detect gaps as large as $20\,$meV at $5\,$K that persist above $9\,$T \cite{Schimmel2023}, as well as ARPES studies, identifying SC surface states around $10\,$K, located at the Fermi arcs  \cite{Kuibarov2024}. Additional, improved resolution ARPES measurements reveal nodes in the superconducting gap of the surface states, hinting towards unconventional, i-wave pairing symmetry\cite{Changdar2025, huang2025}.

At the same time, enhanced bulk superconductivity with $T_\mathrm{c}\approx 2.4\,$K has been reported in related systems, such as Pt$_{1+\delta}$Bi$_2$ with $\delta=0.26$ \cite{Kudo2021}. This compound crystallizes in a trigonal CdI$_2$-type structure (centrosymmetric $P\overline{3}m1$), in which excess Pt intercalates between the PtBi$_2$ layers, removing the Bi-layer corrugation present in $\gamma$-PtBi$_2$. The excess of Pt appears to span a narrow range around $\delta\approx 1/4$ \cite{Kudo2021}, placing this compound close to the Pt$_2$Bi$_3$ composition in the Bi-Pt binary phase diagram \cite{Okamoto1991}. The bulk character of the SC was confirmed by magnetization, $M(T,H)$, and heat capacity measurements, $C_\mathrm{p}(T)$.            

Interestingly, bulk superconductivity with $T_\mathrm{c}\approx 2.4\,$K is again observed when doping $\gamma$-PtBi$_2$ at the Bi-site. A recent report \cite{Takaki2022} shows that a structural transition between the two trigonal modifications, $P31m\to P\overline{3}m1$, occurs as a chalcogen partially substitutes Bi in PtBi$_{2-x}$Ch$_x$ (Ch$\,=\,$Se or Te). The phase transition occurs very close to the PtBi$_2$ composition, needing only $2\%$ ($3\%$) replacement of Bi by Te (Se) in order to induce it. Bulk superconductivity with $T_\mathrm{c}\approx 2.4\,$K emerges in the trigonal centrosymmetric $P\overline{3}m1$ phase, limited to a narrow range of compositions close to the structural transition. Possible origins of this $T_\mathrm{c}$ enhancement have been attributed to an increased density of states compared to $\gamma$-PtBi$_2$, the contribution of soft phonons linked to the structural transition, and/or the suppressed spin splitting of the Fermi surface in the centrosymmetric structure \cite{Takaki2022}.

Until now such studies have been performed on polycrystalline samples only, which may be affected by compositional variations. Furthermore, polycrystals preclude studies of specific crystallographic orientations relative to applied fields. In response to this limitation, we have grown and characterized a series of large crystals of the PtBi$_{2-x}$Ch$_x$ with Ch$\,=\,$Te system, for compositions very close to the reported structural transition. Here, we present a comprehensive study of the physical properties of single crystals with the $x=0.04$ composition ($2\%$ doping), including crystal structural characterization, low-temperature electrical, magnetic, and thermal physical properties, as well as ARPES and point contact measurements.

\section{Experimental Details} 
\label{sec:experimental}

\begin{figure*}
    \centering
    \includegraphics[width=0.98\linewidth]{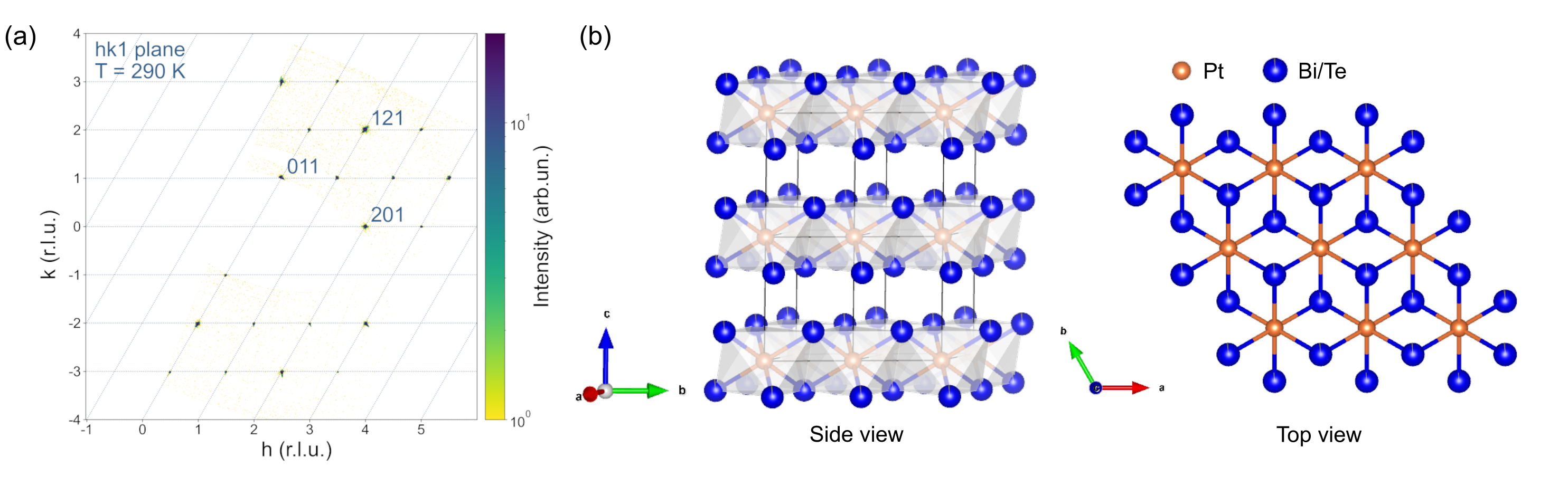}
    \caption{(a) Map of x-ray scattering intensities in the (hk1)-plane, (b) crystal structure of PtBi$_{1.96}$Te$_{0.04}$.}
    \label{fig:XRD}
\end{figure*}

Single crystals of PtBi$_{2-x}$Te$_{x}$ ($x = 0$, $0.02$, $0.03$, $0.04$ and $0.05$) were grown using a stoichiometric-melt-growth method. Stoichiometric amounts of elemental Pt (99.9+ \%, Thermo Scientific), Bi (99.999\%, Alfa Aesar), and Te (99.999\%, Alfa Aesar) powders corresponding to the desired nominal compositions were thoroughly mixed and then loaded and sealed in evacuated quartz ampoules. The sealed ampoules were heated to $800\,^{\circ}$C  at a rate of $100\,^{\circ}$C/h and held at this temperature for $48$ h, before being slowly cooled to $480\,^{\circ}$C at a rate of $2\,^{\circ}$C/h, followed by quenching in cold water. Millimeter-sized crystals with metallic lustre were extracted from the resulting ingot, which readily cleaves leaving mainly flat surfaces with terraces.
The final composition of the crystals was verified by Energy Dispersive Spectroscopy (EDS) at different punctual spots of each sample and by Inductively Coupled Plasma Optical Emission Spectroscopy (ICP-OES). EDS measurements were performed using a scanning electron microscope (Zeiss EVO MA 15) equipped with an energy dispersive X-ray analyser and AZtec software. For the ICP-OES measurements, an iCAP6500 Duo View spectrometer (Thermo Fisher Scientific) was used. 

Single crystal x-ray diffraction (scXRD) measurements were performed at the laboratory facility ``VEGA'', optimized for resolution and sensitivity. This custom-made instrument is equipped with a monochromatized Mo K$_{\alpha}$ radiation source ($\lambda=0.7136\,$\AA) and a CdTe area detector DECTRIS Pilatus with 300000 pixels and no readout noise for high detection efficiency and minimum background. In order to collect the diffraction patterns the $\omega-2\theta$ scan technique was used. Omega sweep step was chosen as $0.5^{\circ}$, while 2$\theta$ was changed from $-36^{\circ}$ to $36^{\circ}$ with $6^{\circ}$ steps. The exposure time for every frame was set to $20\,$s. Recorded diffraction patterns were processed utilizing CrysAlisPro software \cite{CrysAlisPRO}. This included reconstruction of three-dimensional reciprocal lattice, determination of the structure unit cell and peaks’ intensity integration. The structure solution and refinement were carried out using the JANA2006 software package exploiting SuperFlip approach \cite{Jana2006,superflip}.

Magnetization measurements were performed on a MPMS-3 SQUID magnetometer in VSM mode (Quantum Design, QD). Both zero-field cooling (ZFC) and field cooling (FC) protocols were used to collect the data at different applied magnetic fields, both along and perpendicular to the sample ab-plane. Other properties were measured in a Physical Property Measurement System PPMS $9$ Tesla system (QD). Heat-capacity measurements were performed using a heat pulse
relaxation technique. A puck consisting of a frame with a sapphire platform, suspended by 8 thin wires serving as thermal links, was used as a sample holder. The sample was attached to the platform using a small amount of Apiezon N grease. Addenda measurements without the sample were first performed, and then thoroughly subtracted from the measured data.  Electrical resistivity was measured using the corresponding DC four-terminal method. Electrical contacts were made using thin Pt wires of $50\,\mu$m and silver paste. Both the heat capacity and resistivity measurements were performed with the field applied perpendicular to the large flat surface of the samples.  

Pressure-type point contacts (PCs) on a PtBi$_{1.96}$Te$_{0.04}$ single crystal were established directly in a cryostat by mechanically touching thin wires of elemental metals (Cu, Ag and Co) to the $ab$-plane or the edge of the sample. The current–voltage $I(V)$ and differential resistance $dV/dI(V)$ characteristics of PCs were measured by sweeping the dc current $I$ through the PC, on which a small ac current was superimposed, and the first harmonic of the ac signal was detected using a standard lock-in technique in a four-probe configuration. The measurements were performed in the temperature range between $1.5~$K and $10~$K and in magnetic fields up to several Tesla.

Angle-resolved Photoemission Spectroscopy (ARPES) measurements were performed at the SOLARIS synchrotron (URANOS beamline) and IFW-Dresden in mini-ARPES laboratory of German-Ukrainian Research Center for Quantum Materials. The electron energy analyser used at URANOS is hemispherical analyser Scienta DA30L equipped with deflectors. The samples were cleaved \textit{in situ} under a pressure lower than $1 \times 10^{-10}\,\mathrm{mbar}$ and measured at temperatures between 5 and 15 K. Fermi surface maps in Fig.~\ref{fig:ARPES}(a-d) were recorded using FeSuMa electron analyzer \cite{fesuma}. The $\gamma$-PtBi$_2$ reference sample measured with ARPES was grown in a different batch from previously reported $\gamma$-PtBi$_2$ \cite{Kuibarov2024, Changdar2025, Kuibarov2025} and the PtBi$_2$ synthesized in this work. However, this does not affect the results, as $\gamma$-PtBi$_2$ crystals from various batches show very similar Fermi surface maps, as can be seen by comparing Fig.\,\ref{fig:ARPES} with Ref.\,\cite{Kuibarov2024} and Ref.\,\cite{Kuibarov2025}.

Magnetization, heat capacity, resistance, point contact and EDS measurements were all performed on the same Te-doped PtBi$_2$ crystal. Two different crystals of the same batch were chosen to measure scXRD and ARPES, due to size constraints and the need for cleaving.

\section{Results} 
\label{sec:results}

Figure \ref{fig:XRD}-a) shows the hk1) slice of the PtBi$_{1.96}$Te$_{0.04}$ reciprocal lattice, demonstrating correctness of the found unit cell (other slices are also presented in the Supplementary \cite{supp}). Although the simple centric test for the entire peak set was not reliable, the structure solution and following refinement converged directly to the centrosymmetric model. Thus, the crystal is indexed to the $P\overline{3}m1$ space group, consistent with previous reports that pointed out the suppression of the distortion in the Pt triangular network present in the pristine PtBi$_{2}$, stabilizing the centrosymmetric structure \cite{Takaki2022}. The obtained lattice parameters are $a=b=4.0545(10)\,$\AA\ and $c=5.620(18)\,$\AA, with an additional refined internal parameter $z/c=0.254(11)$ describing the position of the (Bi$_{0.98}$Te$_{0.02}$) atoms. In Fig.~\ref{fig:XRD}-b) we present the refined lattice structure, characterized by layers of Pt(Bi$_{0.98}$Te$_{0.02}$)$_2$ stacked along the c-axis. Within each layer, Pt atoms are coordinated by six (Bi$_{0.98}$Te$_{0.02}$) atoms, forming Pt(Bi$_{0.98}$Te$_{0.02}$)$_6$ octahedra. These octahedra are linked through shared edges, resulting in the Pt(Bi$_{0.98}$Te$_{0.02}$)$_2$ layers where the Pt atoms create a triangular network.

The chemical composition analysis performed by EDS on cleaved surfaces of a crystal gives relative ratios for the three elements (Pt:Bi:Te) = (37.45:61.33:1.22) at\%, indicating a deficiency in the total amount of (Bi+Te) with respect to Pt. This $p$-element deficiency has been consistently observed for crystals with other $x$ compositions, with the exception of stoichiometric $x=0$ samples with (Pt:Bi) = (33.4:66.6) at\%. To corroborate this result, we also performed ICP-OES measurements on another crystal from the same batch. This study gives a ratio (Pt:Bi:Te) = (35.6:63.0:1.4) at\%, indicating a consistent but smaller deficiency ratio $\mathrm{(Bi+Te):Pt}\approx 1.81$. Note, however, that the measured $\mathrm{Te:Bi}\sim 0.02$ corresponds to the expected ratio, so we will keep the nominal $x$ composition when referring to the current samples in the following.

\begin{figure}
    \centering
    \includegraphics[width=0.8\linewidth]{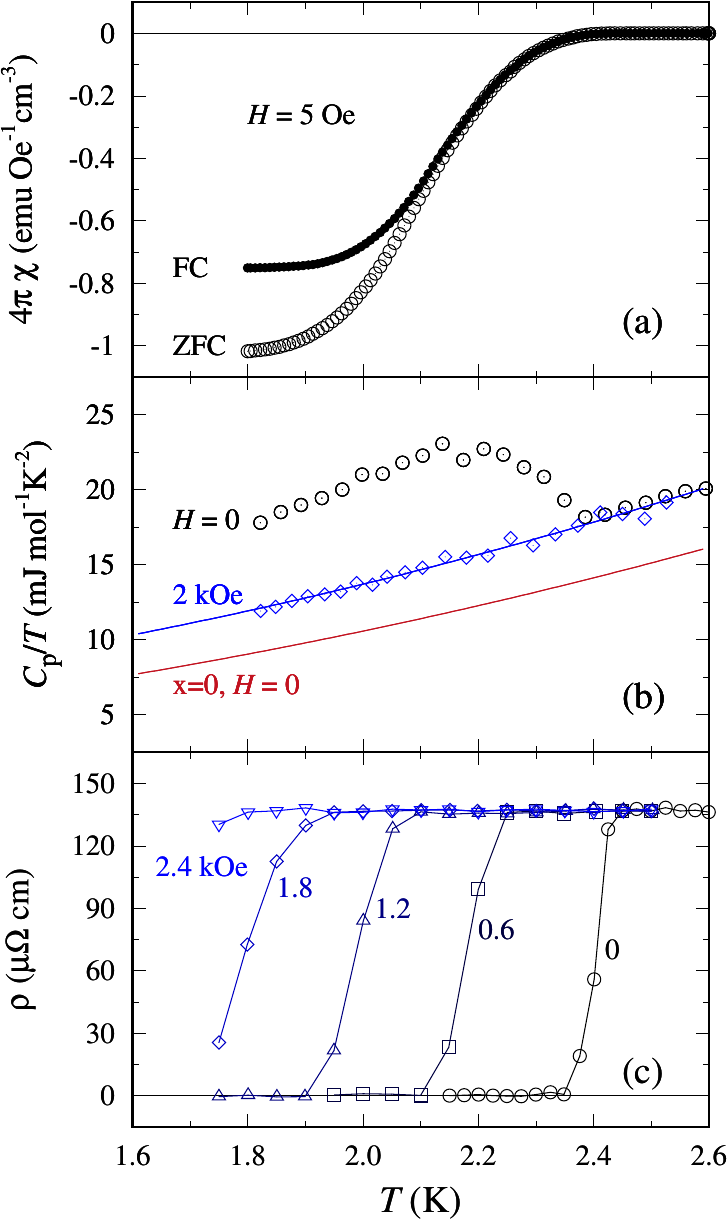}
\caption{Bulk superconducting transition in PtBi$_{1.96}$Te$_{0.04}$ at $T_\mathrm{c} \approx 2.4~$K, as seen from (a) magnetization measurements, (b) heat capacity and (c) electrical resistivity measurements. Panel (a) displays the dc magnetic susceptibility measured at $H=5~$Oe following ZFC and FC protocols. Panel (b) depicts our $H=0$ and $2~$kOe data, a field high enough to suppress any evidence of SC in the measured $T$-range. Full lines are fittings to the $H=2~$kOe data (blue line) and $H=0$ data for our $x=0$ reference sample. Panel (c) shows the resistivity SC transition at different applied fields, reaching a large normal state value $\rho(2.5\,\mathrm{K})\approx 135\,\mu\Omega\,\mathrm{cm} \sim\rho(300\,\mathrm{K})$.}
    \label{fig:properties}
\end{figure}

Figure \ref{fig:properties} displays a collection of results intended to establish and characterize the bulk nature of superconductivity for our sample with $x=0.04$ Te-doping. The upper panel (a) depicts the dc magnetic susceptibility $\chi\equiv M/H$ measured at $H=5\,$Oe with field applied along the $ab$-plane, showing a strong diamagnetic response below $T\sim 2.4\,$K. We chose a representation that highlights the full shielding observed at low temperatures, where the ZFC curve reaches $4\pi\chi\approx -1$ as expected for SC in magnetic fields lower than the critical field $H_\mathrm{c1}$, in contrast to the other compositions, where a reduced superconducting volume fraction was observed (see Supplemental Material \cite{supp}). The FC curve, on the other hand, saturates around $4\pi\chi(1.8\,\mathrm{K})\approx -0.75$, probably indicative of a weak pinning effect. We take $1$\% of the full shielding value to define the onset temperature from our magnetization measurement, $T_{\mathrm{c}}^{\mathrm{on}}=2.37\,$K.

The low-$T$ heat capacity of PtBi$_{1.96}$Te$_{0.04}$ is presented in Fig.~\ref{fig:properties}-b) in a $C_\mathrm{p}/T$ versus $T$ representation, both for zero field and $H=2\,$kOe. As a further reference, notice the red curve in Fig.~\ref{fig:properties}-b) that represents the fit to the $\gamma$-PtBi$_2$, $x=0$ data, $C_\mathrm{p}/T[\mathrm{mJ}/\mathrm{mol\,K}^2]=2.5(6)+2.0(2) T^2$. The zero-field curve displays a distinctive anomaly associated with the second-order SC transition, from which we estimate $T_\mathrm{c}=2.3\,$K from its midpoint. A magnetic field of $2\,$ kOe completely removes the heat capacity anomaly, leaving only normal electrons to contribute to $C_\mathrm{p}$ in the studied temperature range. These data can be described as $C_\mathrm{p}/T=\gamma + \beta T^2$, where $\gamma=4.3(3)\,\mathrm{mJ}/\mathrm{mol\,K}^2$ is the Sommerfeld coefficient and $\beta= 2.35(5)\,\mathrm{mJ}/\mathrm{mol\,K}^4$ accounts for the low temperature phonon contribution (see the full line in the figure). The resulting Debye temperature $\Theta_{\mathrm D}\sim 135\,$K can be estimated from $\beta=1944 r/\Theta_{\mathrm D}^3\,\mathrm{mJ}/\mathrm{mol\,K}^4$ \cite{Gopal1966}, where $r=3$ is the number of atoms per formula unit. Both the Sommerfeld coefficient and the resulting Debye temperature compare well with the values found for PtBi$_{1.8}$Te$_{0.2}$ with $\gamma = 4\,\mathrm{mJ}/\mathrm{mol\,K}^2$ and $\Theta_{\mathrm D}\sim 135\,$K \cite{Takaki2022}. From an estimation of the heat capacity jump around $T_{\mathrm{c}}$ we get $ \Delta C / \gamma T_{\mathrm{c}} \approx  2.1$, hinting towards strong coupling superconductivity \cite{TinkhamSC}. Both, the larger $\gamma$ and specific heat jump at $T_\mathrm{c}$, can be interpreted as further indication of a large superconducting volume fraction involving the bulk of the samples.

In Fig.~\ref{fig:properties}-c) we show the resistive SC transition at fields between $0$ and $2.4\,$kOe. The normal state resistivity is high, $\rho_{\mathrm{N}}\approx 135\,\mu\Omega\,$cm, and close to the room temperature value, giving a residual resistivity ratio $RRR\equiv \rho(300\,\mathrm{K})/\rho_{\mathrm{N}}\approx 1.1$. At zero field, the transition is sharp. We use a $50\,\%$ criterion to define $T_{\mathrm{c}}^\rho=2.40\,$K for $\rho(T_{\mathrm{c}}^\rho)\equiv 0.5\rho_{\mathrm{N}}$. The transition broadens and shifts to lower temperatures as the applied magnetic field increases, only the onset remaining at $2.4\,$kOe in the studied $T$-range. $T_{\mathrm{c}}(H)$ decreases roughly linearly with $H$.

\begin{figure}
    \centering
    \includegraphics[width=0.95\linewidth]{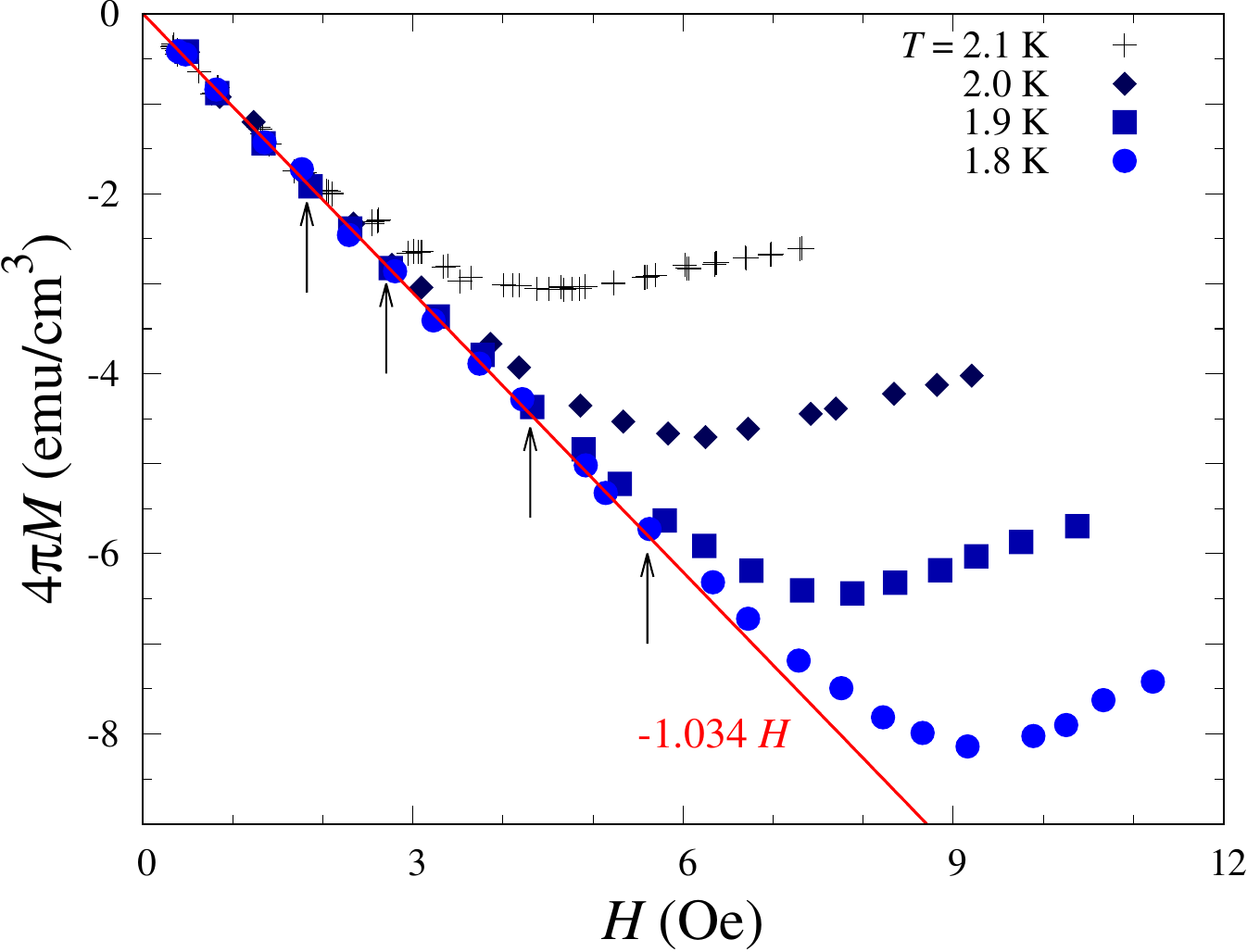}
    \caption{Low field magnetization measurements of PtBi$_{1.96}$Te$_{0.04}$ for different constant temperatures between $1.8\,$K and $2.1\,$K. Black arrows indicate the deviation of the measured data from the linear regime, serving as estimation for the penetration field $H_{\mathrm{P}}\sim H^{\parallel}_\mathrm{c1}$.}
    \label{fig:hc1}
\end{figure}

\begin{figure}
    \centering
    \includegraphics[width=0.95\linewidth]{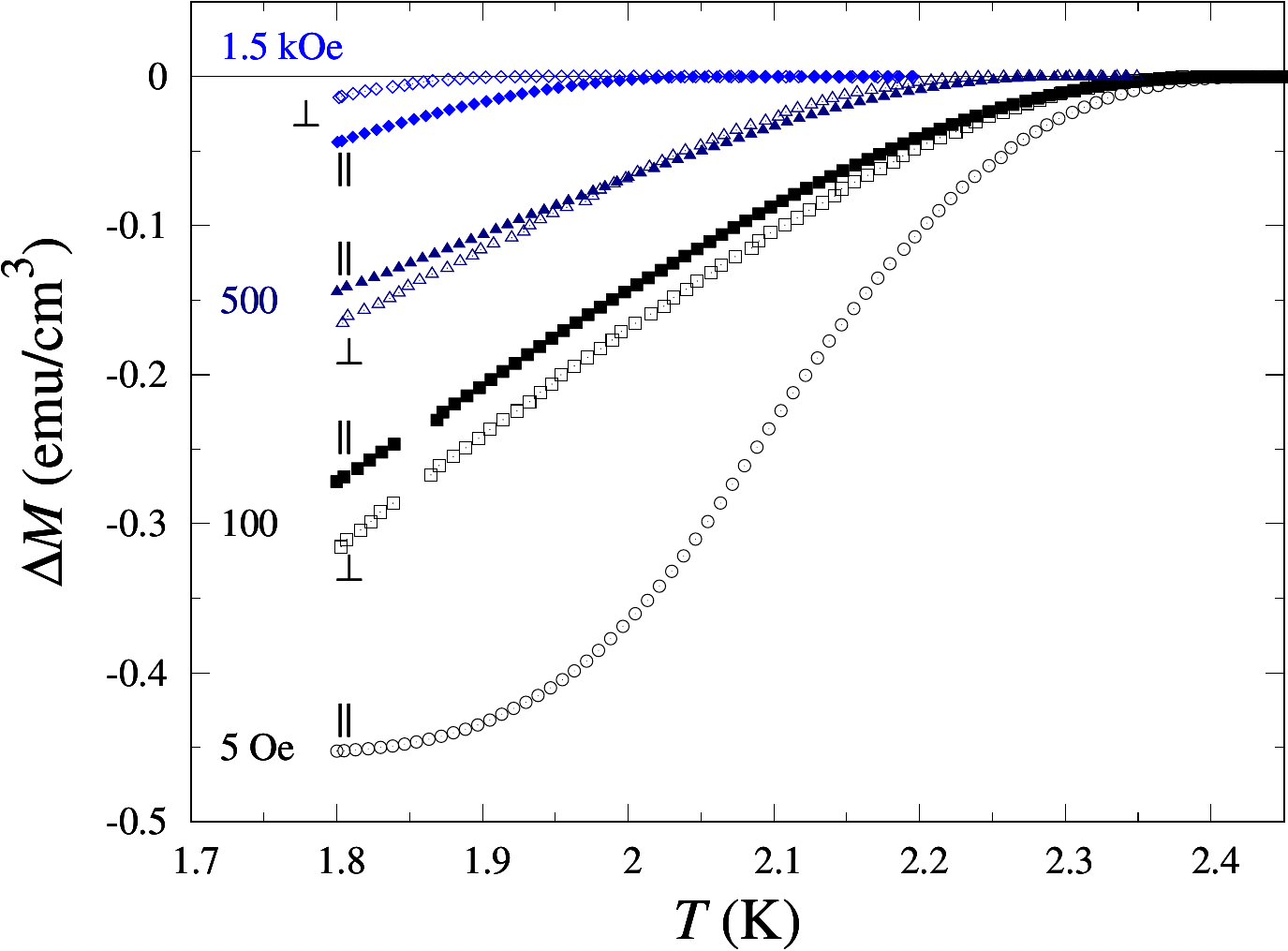}
    \caption{Temperature dependent magnetization of PtBi$_{1.96}$Te$_{0.04}$ at constant magnetic fields between $0$ and $1.5~$kOe. The data are presented as $\Delta M\equiv M(T)-M(T\gtrsim T_{\mathrm{c}})$ to account for the weak magnetization in the normal state. Magnetic fields were applied both parallel ($\parallel$) and perpendicular ($\perp$) to the $ab$-plane.}
    \label{fig:hc2}
\end{figure}

\begin{figure}
    \centering
    \includegraphics[width=0.95\linewidth]{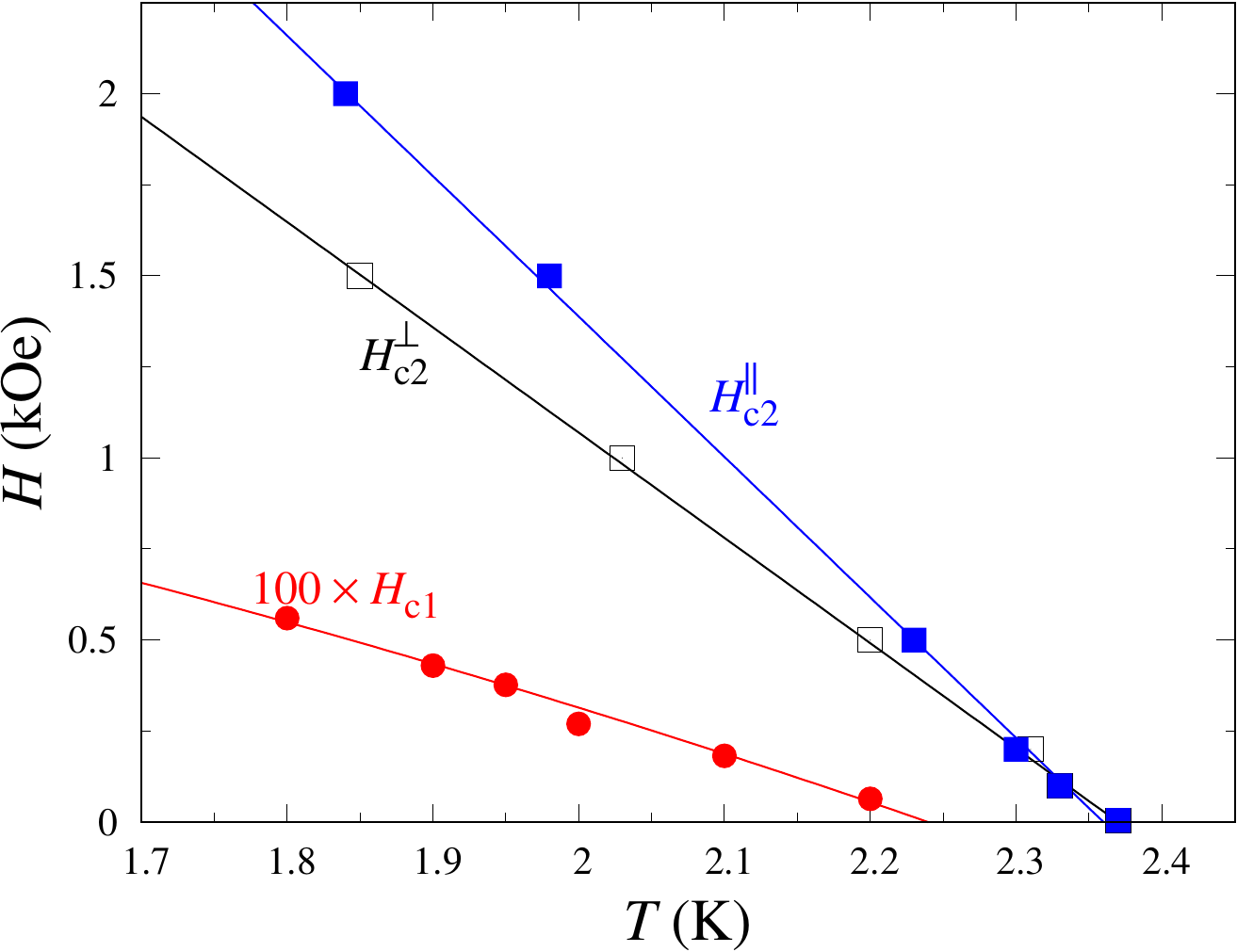}
    \caption{Lower critical field $H_{\mathrm{c}1}^{\parallel}$ (data from Fig.~\ref{fig:hc1}) and upper critical fields $H_{\mathrm{c}2}$ for magnetic fields applied parallel and perpendicular to the $ab$-plane. The corresponding fits (continuous lines) are described in the text. Notice that the $H_{\mathrm{c}1}$ data is multiplied by $100$.}
    \label{fig:phase_diagram}
\end{figure}

The low-field magnetization is analyzed in more detail in Fig.~\ref{fig:hc1}, showing $4\pi M(H)$ measured at four different fixed temperatures. As expected, the data initially follow a linear dependence, $M\sim -H$, up to a field at which the curve deviates upward (see the arrows), developing a minimum in $M(H)$ at a field approximately twice as large. We associate this initial deviation from a straight line with the first penetration field $H_{\mathrm{P}}$, at which vortices begin to penetrate the bulk of the sample. The red straight line shows the fitted Meissner dependence corresponding to $4\pi\chi\approx -1$, as discussed in the context of Fig.~\ref{fig:properties}-a. An absolute slope slightly larger than $1$ is probably associated with a small demagnetizing factor of the flat sample oriented along the applied magnetic field ($H\parallel ab$-plane). This is indicative that in this particular configuration, we can approximate $H_{\mathrm{P}}\approx H^{\parallel}_\mathrm{c1}$, the lower critical field (see e.g.~\cite{Klein2010}). The evolution of $H^{\parallel}_{\mathrm{c1}}(T)$ is shown in Fig.~\ref{fig:phase_diagram}, multiplied by a factor $100$ to make the data visible on the selected scale. The full line corresponds to a fit of our data to the expression $H^{\parallel}_{\mathrm{c1}}(T)=H^{\parallel}_{\mathrm{c1}}(0)\left[1-(\frac{T}{T_\mathrm{c1}})^2\right]$, with parameters $H^{\parallel}_{\mathrm{c1}}(0)=15.5(7)\,$Oe and $T_\mathrm{c1}=2.24(2)\,$K.

Magnetization measurements at higher constant fields are shown in Fig.~\ref{fig:hc2}, both for magnetic fields applied along the $ab$-plane ($\parallel$) and perpendicular to it ($\perp$). We plotted the data for a given $H$ as $\Delta M\equiv M(T)-M(T\gtrsim T_{\mathrm{c}})$, and defined $T_{\mathrm{c}}^{\mathrm{on}}$ as the temperature, at which the magnetization measured at $H=5~$Oe reaches 1\% of its lowest value $\Delta M(5\,{\mathrm Oe},1.8\,{\mathrm K})$. Demagnetizing effects have not been considered in the calculation of $H_{c2}$, due to the small values of Magnetization. The resulting values $T_{\mathrm{c}}^{\mathrm{on}}(H)$ are shown in Fig.~\ref{fig:phase_diagram}, along with fitting curves of the form $H_{\mathrm{c2}}(T)=h_{\mathrm{c2}}\left(T-T_\mathrm{c2}\right)$. The resulting fitting parameters are $h^{\parallel}_{\mathrm{c2}}=3.87(7)\,\mathrm{kOe}/$K and $h^{\perp}_{\mathrm{c2}}=2.89(4)\,\mathrm{kOe}/$K, with $T_\mathrm{c2}=2.36(2)\,$K. Taking into account the Werthamer-Helfand-Hohenberg formula $H_{\mathrm{c2}}(0)=0.693\cdot T_\mathrm{c2} h_{\mathrm{c2}}$ \cite{Werthamer1966}, we estimate $H^{\parallel}_{\mathrm{c2}}(0)\approx 6.3\,$kOe and $H^{\perp}_{\mathrm{c2}}(0)\approx 4.7\,$kOe, giving a critical field anisotropy ratio $\Gamma_{2}\equiv H^{\parallel}_{\mathrm{c2}}/H^{\perp}_{\mathrm{c2}} \sim 1.3$. 

We can use the critical field values $H_{\mathrm{c1}}$ and $H_{\mathrm{c2}}$ to estimate the superconducting coherence length, $\xi$, and penetration depth, $\lambda$. For anisotropic (layered) superconductors, $H^{\perp}_{\mathrm{c2}}(0) = \frac{\Phi_0}{2 \pi \xi_{ab}^{2}}$ and $H^{\parallel}_{\mathrm{c2}}(0) = \frac{\Phi_0}{2 \pi \xi_{ab}\xi_{c}}$, where $\Phi_0=2.07\times 10^{-7}\,\mathrm{G}\,\mathrm{cm}^2$ is the magnetic flux quantum \cite{TinkhamSC}. The resulting coherence lengths are $\xi_{ab} \approx 26\,$nm, and $\xi_c \approx 20\,$nm. We further assume $\Gamma_{1}\equiv H^{\perp}_{\mathrm{c1}}/H^{\parallel}_{\mathrm{c1}}\sim \Gamma_{2}$ \cite{TinkhamSC}, and use the expression $\Gamma_1 H^{\parallel}_{\mathrm{c1}}(0)\approx \frac{\Phi_0}{4 \pi \lambda_{ab}^2}\left[ \ln(\kappa)+0.5 \right]$ to estimate $\lambda_{ab}\approx 540\,$nm. Here, $\kappa=\lambda_{ab}/\xi_{ab}\approx 20$ is the dimensionless Ginzburg-Landau parameter.

\begin{figure}
    \centering
    \includegraphics[width=0.98\linewidth]{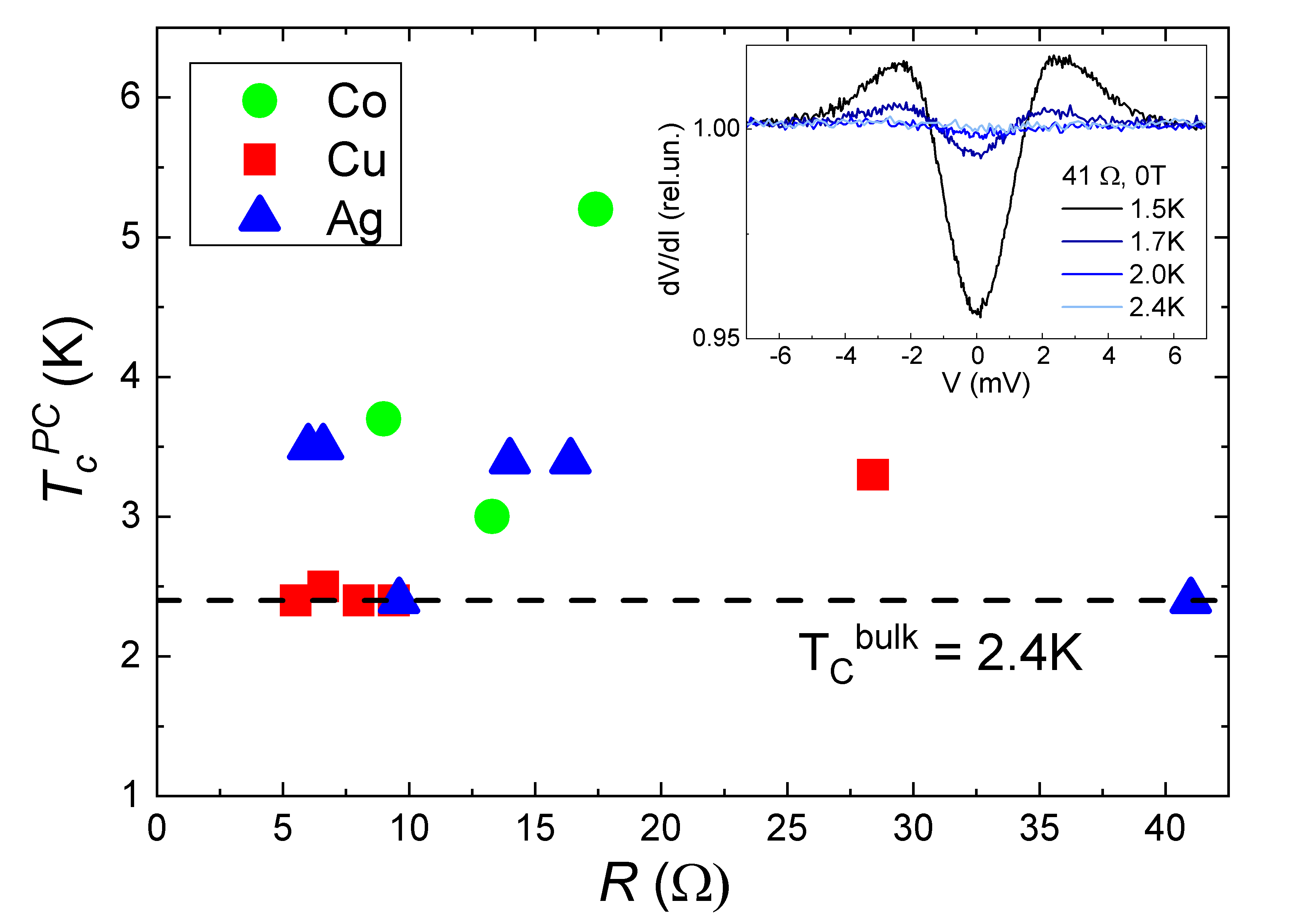}
    \caption{$T_{\mathrm{c}}$ distribution for all measured heterocontacts with normal (Ag, Cu) and ferromagnetic (Co) metal counterelectrodes. The abscissa shows PC resistance in the normal state. Inset: temperature variation of $dV/dI$ of PtBi$_{1.96}$Te$_{0.04}$-Ag PC with $T_{\mathrm{c}}\approx 2.4\ $K. Normal state resistance is $R=41 \, \Omega$.}
    \label{fig:PCS}
\end{figure}

We also studied point contacts on PtBi$_{1.96}$Te$_{0.04}$ using different metals, to gain further insight into the superconducting properties of our crystals. The resuls of these PC measurements are summarized in Fig.~\ref{fig:PCS}. The inset of the figure displays representative differential resistance curves, $dV/dI$, for a PtBi$_{1.96}$Te$_{0.04}$-Ag PC with $T_{\mathrm{c}} \approx 2.4\,$K, which is the same as the value for the bulk material. The shape of $dV/dI$, with a deep zero-bias minimum and symmetrical side maxima, is typical for all the measured contacts and has similarities to measurements in $\gamma$-PtBi$_2$ \cite{Bashlakov2022}. The zero-bias minimum in $dV/dI$ is connected to the superconducting transition, the intensity of which decreases both by increasing temperature or applied magnetic field. Sharp side peaks with a featureless zero-bias dip are characteristic of the nonspectral regime in the PC. In this regime the superconducting state is suppressed by high current density, that increases with voltage and results in the temperature rise of the contact region. As a consequence, our data can only provide information about the critical temperature and critical magnetic field values of superconductivity at the PCs.

The main panel of Fig.~\ref{fig:PCS} demonstrates the distribution of $T_{\mathrm{c}}$ for all measured heterocontacts based on PtBi$_{1.96}$Te$_{0.04}$ with normal (Cu, Ag) and ferromagnetic (Co) metal counterelectrodes. The average superconducting critical temperature for all measured PCs amounts to $\overline{T_{\mathrm{c}}}=3.1\,$K, exceeding the bulk value but overlapping within one standard deviation $\sigma_{PC}= 0.8\,$K. The largest $T_{\mathrm{c}}$ in the PCs reaches about $5\,$K, which is approximately twice the value for the bulk, and was observed in a PC with a magnetic Co tip. The critical magnetic field is also significantly enhanced, reaching up to $4\,$T as in the case of PCs on pure $\gamma$-PtBi$_{2}$ \cite{Bashlakov2022}.

\begin{figure*}
    \centering
    \includegraphics[width=1\linewidth]{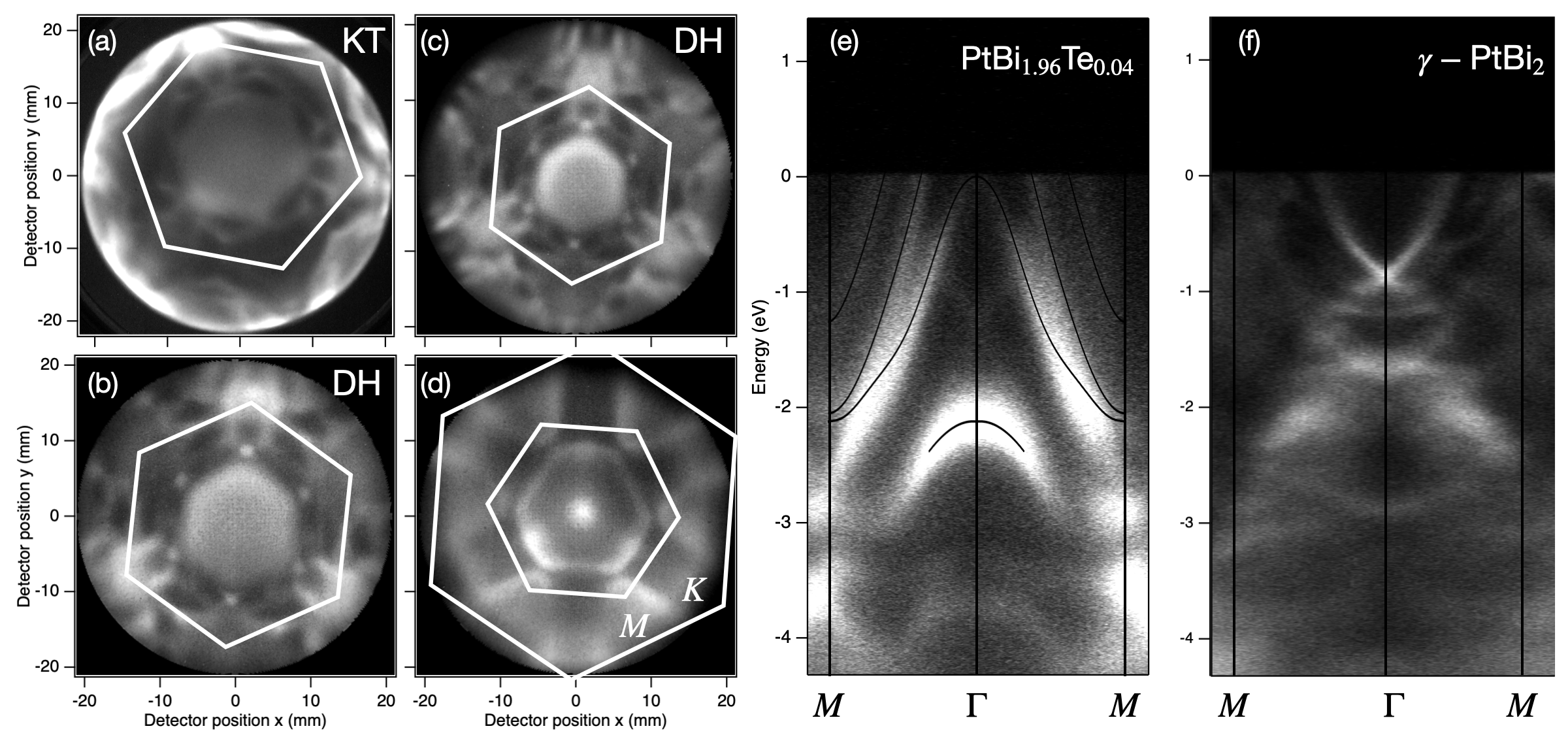}
    \caption{ARPES characterization of $\gamma$-PtBi$_2$ and  PtBi$_{1.96}$Te$_{0.04}$. (a,b) Fermi surface maps of $\gamma$-PtBi$_2$ for the Kagome-type (KT) and decorated honeycomb (DH) terminations, respectively, measured using $21.2\,$eV photons and a $60\,$V accelerating potential. (c) Fermi surface maps of $\gamma$-PtBi$_2$ DH termination,  measured using $21.2\,$eV photons and a $90\,$V accelerating potential.  (d) Fermi surface map of PtBi$_{1.96}$Te$_{0.04}$ measured with $21.2\,$eV photons and $90\,$V accelerating potential; the inner and outer hexagons represent the Brillouin zones for the $\gamma$-PtBi$_2$ and PtBi$_{1.96}$Te$_{0.04}$ phases, respectively. (e) Energy-momentum intensity distribution of PtBi$_{1.96}$Te$_{0.04}$ along the $M-\Gamma-M$ direction measured with $120\,$eV photon energy. The solid black lines represent the nonfolded bands from band structure calculations with SOC adapted from Ref.~\cite{Palumbo2025}. A rigid energy shift of $\sim 0.53$\,eV was applied to the calculated bands. (f) Energy-momentum intensity distribution of $\gamma$-PtBi$_2$ along the $M-\Gamma-M$ direction measured at the same experimental condition as (e).}
    \label{fig:ARPES}
\end{figure*}

The band structure of pure $\gamma$-PtBi$_2$ and PtBi$_{1.96}$Te$_{0.04}$ exhibit profound differences driven by their underlying crystal symmetries. As observed from Figs.~\ref{fig:ARPES}-a) and b) the Fermi surfaces of $\gamma$-PtBi$_2$ depend heavily on the exposed surface termination, because of its non-centrosymmetric nature, while the Fermi surface of the Te-doped compound, presented in Fig.~\ref{fig:ARPES}-d), shows a single stable termination. This is a consequence of the substitution of Bi with Te, suppressing the non-centrosymmetric crystalline distortion inherent to the parent $\gamma$-PtBi$_2$ phase.
Since the non-centrosymmetric unit cell is a supercell of the centrosymmetric phase, the Brillouin zone of the Te-doped compound is significantly larger \cite{Palumbo2025}, which can be observed by comparing the Fermi surfaces measured under the same conditions in Figs.~\ref{fig:ARPES}-c,d).

Finally, Fig.~\ref{fig:ARPES}-e) illustrates the excellent quantitative agreement between the experimental band dispersion of PtBi$_{1.96}$Te$_{0.04}$ and the calculated band structure for the centrosymmetric reference phase~\cite{Palumbo2025}. This close correspondence between experiment and DFT confirms that the breaking of inversion symmetry drives the metal-to-semimetal transition in PtBi$_2$. Consequently, due to the inversion-symmetric nature of $\text{PtBi}_{1.96}\text{Te}_{0.04}$, Fermi arcs are neither theoretically expected nor experimentally observed in the Te-doped compound. To further highlight the changes between the undoped and doped PtBi$_2$ compound, we also show the markedly different energy-momentum intensity distribution of $\gamma$-PtBi$_2$, measured under the same experimental conditions as Te-doped PtBi$_2$, in Fig.~\ref{fig:ARPES}-f).


\section{Discussion}

The analysis of the data  presented in the previous section leads us to conclude that PtBi$_{1.96}$Te$_{0.04}$ is a type II superconductor within the strong coupling limit. Magnetization measurements on single crystals clearly show that the observed superconductivity originates from the bulk, with a full shielding in ZFC measurements, and allow us to estimate a low value for the critical field anisotropy, $\Gamma\sim 1.3$. Additionally, the heat capacity displays a jump of large magnitude at the transition temperature $T_{\mathrm{c}}$, with $\Delta C / \gamma T_{\mathrm{c}} \approx  2.1$, larger than $1.43$, the expected ratio for a BCS weak-coupling superconductor.

The single crystal X-ray diffraction patterns, here reported for the first time, confirm the $P\overline{3}m1$ crystal structure for the low $x=0.04$ Te-doping. Our findings on single crystals are thus consistent with a previous report on PtBi$_{2-x}$Te$_{x}$ that studied polycrystalline samples \cite{Takaki2022}. However, unlike that report, we find that our $x=0.04$ crystals show an atomic concentration ratio $(\mathrm{Bi}+\mathrm{Te}):\mathrm{Pt}\approx 1.8$, lower than the expected nominal value. This can either originate from a Pt excess in the crystals, a $p$-element deficiency, or a combination of both. The centrosymmetric trigonal $P \overline{3}m1$ structure can accommodate excess Pt as an intercalant between PtBi$_2$ layers, resulting in a well defined ratio $\mathrm{Bi}:\mathrm{Pt}=2:(1+\delta)=1.59$ \cite{Kudo2021}, smaller than the one we measured. Alternatively, Bi deficient PtBi$_{2-\tilde{\delta}}$ crystals have also been reported, with $\tilde{\delta}\approx 0.4$ \cite{Xing2020}. This again results in  $\mathrm{Bi}:\mathrm{Pt}=(2-\tilde{\delta}):1\approx 1.6$, although in this case the $P31m$ structure is preserved \cite{Xing2020}. We attempted to refine the XRD data by considering both EDS and ICP-OES relative occupancies, however, no corresponding improvement in the goodness-of-fit was obtained. Due to this uncertainty of the stoichiometry, all parameters derived from the molar mass of the formula unit inherit an error, which we assess here by determining the scenarios of complete Pt-excess or complete Bi-deficiency. Hence, we get $\gamma=4.33 \pm 0.28\,\mathrm{mJ}/\mathrm{mol\,K}^2$ and $\beta= 2.35 \pm 0.15 \,\mathrm{mJ}/\mathrm{mol\,K}^4 $. Other synthesized crystals of PtBi$_{2-x}$Te$_x$ with nominal concentrations $x = 0, 0.02, 0.03, 0.05$ show a relative (Bi+Te) deficiency that grows systematically with $x$ in the $\sim 1.94-1.79$ range (ICP-OES values). In these crystals the superconducting volume fraction at $1.8\,$K increases with $x$, starting from roughly 60\% at $x=0.02$ (see Supplemental Material \cite{supp}). Since the relative deficiency of (Bi+Te) correlates with the amount of Te-doping, we also cannot exclude the possibility that this (Bi+Te)-deficiency plays a role in the observed superconducting and structural transition.


A broadening of the SC transition, as observed in our specific heat data, is usually interpreted as an effect of non-homogeneous composition across the sample. Nevertheless, the critical temperature $T_\mathrm{c}(x)$ in our crystals is only weakly affected by the Te-doping level $x$ \cite{supp}, consistent with the findings in Ref.~\cite{Takaki2022}. This suggests that Te-inhomogeneity cannot be the sole reason for the transition broadening \footnote{Punctual EDS analysis on more than 20 locations of a   $x=0.04$ crystal provide a mean 2\% substitution of Bi by Te, with a standard deviation $\sigma_{\mathrm{Te}}\approx 0.5$\%. The critical temperature decreases with Te doping at a rate $T'_{\mathrm{c}}\lesssim -0.02~\mathrm{K}/$\%Te \cite{Takaki2022}, while we estimate $T'_{\mathrm{c}}\sim -0.013~\mathrm{K}/$\%Te for our crystals with $x\leq 0.05$. Thus, within one standard deviation ($\sim 70$\% of the volume of the sample), $\Delta T_\mathrm{c}(\pm \sigma_{\mathrm{Te}}) \lesssim 0.02~$K, while the $C_{\mathrm{P}}(T)$ broadening reaches roughly $5$ times that value.}, and that an additional mechanism such as remaining non-uniform strain in the crystals is playing a role. Both inhomogeneities and strain have recently been claimed to lead to particular magnetotransport results in SC single crystal flakes of the related PtBi$_{2-y}$Se$_y$ system \cite{Samukawa2025}, while strain due to thermal treatment has been found to induce filamentary superconductivity in wrinkled thin flakes of $P\overline{3}m1$ PtSe$_2$ , again at $T_\mathrm{c}\sim 2.4~$K \cite{Yuan2021}. In this context, a potential effect of local tensions has also been considered to be relevant in $\gamma$-PtBi$_2$ as a mechanism that could enhance the measured $T_{\mathrm{c}}$ \cite{Zabala2024} to values twice as large as the $T_{\mathrm{c}}\sim 0.6~$K reported in Ref.~\cite{Shipunov2020}. In a limiting case, tension could play a role in locally inducing the $P31m \to P\overline{3}m1$ transformation due to the strong contraction of the $c/a$ ratio across the structural transition \cite{Takaki2022}. This possibility should be examined in detail whenever type II superconductivity with a $T_{\mathrm{c}}$ of about $2.4~$K is observed in studies of pristine PtBi$_2$.

The $P31m \to P\overline{3}m1$ structural transition was reproduced by band structure calculations, assuming a volume change due to the differing atomic radii of Bi and Te and a corresponding $x$-increase in the electron count \cite{Vocaturo2024}. Despite this minute electron doping to the parent $\gamma$-PtBi$_2$ compound, a qualitative difference between the normal states of both systems is clearly reflected in the absence of Fermi arcs in the $x = 0.04$ sample, in contrast to the case of high-quality $\gamma$-PtBi$_2$ single crystals, where Fermi arcs are detected for both surface terminations (corrugated and non-corrugated) of the non-centrosymmetric structure \cite{Kuibarov2024}. The suppression of Fermi arcs is to be expected in a non-magnetic system as inversion symmetry is restored. Along with this, additional symmetries are expected to lead to a semimetal-to-metal transition \cite{Palumbo2025}, with a corresponding increase in the density of states at the Fermi level, manifested in the approximate doubling of the Sommerfeld coefficient measured for $x = 0.04$ with respect to the $x = 0$ sample. The semimetal-to-metal transition is also accompanied by substantial changes in the band structure, as reported by Ref.\,\cite{Palumbo2025}. This is consistent with our ARPES measurements, showing excellent quantitative agreement between calculations and experiment.

Point contact experiments with low contact resistances using Cu-tips show consistently a critical temperature that corresponds to the bulk, $T_\mathrm{c}\approx 2.4\,$K, while Ag-tips display an increased $T^\mathrm{PC}_\mathrm{c}\approx 3.5\,$K. PCs creation typically involves the application of local mechanical pressure. Such pressure could, in principle, induce local strain and thereby influence the measured $T_\mathrm{c}$. However, to the best of our knowledge, no pressure-dependent studies on Te-doped samples have been reported to clarify this effect. Another origin for the enhanced $T_c$ could be a marginal intergrowth phase in the PtBi$_{1.96}$Te$_{0.04}$ crystals due to their vicinity to the phase transition to the non-centrosymmetric, trigonal structure of pure PtBi$_2$ \cite{Takaki2022}. From this perspective, the presence of some non-centrosymmetric phase with increased-$T_c$ surface superconducting states in the contact area \cite{Schimmel2023, Kuibarov2024}, could be a reason for the observed $T_c$ enhancement \cite{Bashlakov2022}. On the other hand, measurements with Ag and Cu tips show only moderate ($T^\mathrm{PC}_\mathrm{c}/T_\mathrm{c}\sim 1.5$) or no enhancement in our experiment, while in the case of $\gamma$-PtBi$_2$ an order of magnitude increase in $T^\mathrm{PC}_\mathrm{c}$ has been repeatedly observed \cite{Kvitnitskaya2025}, with $T^\mathrm{PC}_\mathrm{c}>6\,$K. PC studies on a number of materials such as FeSe \cite{Naidyuk2016}, AFe$_2$As$_2$ \cite{Naidyuk2020} or MoTe$_2$ \cite{Naidyuk2018} demonstrated local enhancement of $T^\mathrm{PC}_\mathrm{c}$ that do not seem to be related to plain pressure effects. One of these mechanisms could be at play in PtBi$_{1.96}$Te$_{0.04}$, resulting in an enhancement of both $T^\mathrm{PC}_\mathrm{c}$ and $H^\mathrm{PC}_\mathrm{c}$ for non-magnetic tips. Note that the confirmation of an eventual enhancement of T$_c$ in experiments using a (magnetic) Co-tip requires more statistics, a matter that is beyond the scope of this paper.

In summary, our PtBi$_{1.96}$Te$_{0.04}$ data in single crystals confirm \cite{Takaki2022} the drastic changes observed in trigonal PtBi$_2$ upon substituting just $2$\% of Bi with Te: conventional bulk superconductivity appears in its centrosymmetric phase at $T_\mathrm{c}\approx 2.4~$K. In this phase, as expected, our ARPES measurements confirm the absence of Fermi arcs, one of the key observations that support the unconventional surface superconducting state reported in $\gamma$-PtBi$_2$ at elevated temperatures \cite{Schimmel2023,Kuibarov2024,Changdar2025}. As hinted in Ref.~\cite{Vocaturo2024}, both pressure and Te-doping should result in the Weyl nodes of $\gamma$-PtBi$_2$ approaching the Fermi level, possibly enhancing their experimental accessibility despite the eventual substitutional disorder induced by doping. This calls for further investigations on samples with finely controlled doping level, particularly in the low-doping regime.  

\section{Acknowledgments}
This work was supported by the Deutsche Forschungsgemeinschaft (DFG, German Research Foundation) through the Würzburg–Dresden Cluster of Excellence ‘Complexity and Topology in Quantum Matter’ – ct.qmat (EXC 2147, project ID 390858490) and in the Collaborative Research Center ‘Correlated Magnetism: From Frustration to Topology’ (SFB 1143, project ID 247310070). S.C., Al.K., V.B. and B.B. were supported by the Bundesministerium für Forschung, Technologie und Raumfahrt (BMFTR) funding through project No. 01DK240008 (GU-QuMat). Research at the National Synchrotron Radiation Centre SOLARIS is supported by the Ministry of Science and Higher Education, Poland, under contract no. 1/SOL/2021/2. O.K. acknowledges the DFG Grant (BU 887/31-1) and the support by the National Academy of Sciences of Ukraine under project $\Phi$19-5, while Al.K., V.B. acknowledge the support by the National Academy of Sciences of Ukraine under the "Looking for topological superconductivity in cuprates and iron-based superconductors" grant. The authors would like to thank Andrea Voss for performing ICP-OES measurements and Sebastian Gaß for technical assistance during the measurements. We are indebted to Sabine Wurmehl and Yurii G.~Naidyuk for fruitful discussions. P.P. benefited from discussions with Jorge Facio and Pablo Cornaglia.

\bibliography{PtBi2-refs3}

@article{Alekseevski1953,
  title = {Invstigations of Bismuth alloys at very low temperatures},
  author = {Alekseevskii,N.E. and Gaidukov,Yu.P. },
  journal = {J. Exptl. Theoret. Phys. (U.S.S.R.)},
  volume = {25},
  pages = {383},
  year = {1953},
  doi = {}}

@article{Bashlakov2022,
  title = {Electron-phonon interaction and point contact enhanced superconductivity in trigonal {PtBi$_2$}},
  author = {Bashlakov,D.L. and Kvitnitskaya,O.E. and Shipunov,G. and Aswartham,S. and ~Feya, O.D. and ~Efremov,D.V. and ~Büchner,B. and ~Naidyuk,Yu.G.},
  journal = {Low Temp. Phys.},
  volume = {48},
  pages = {747},
  year = {2022},
  doi = {10.1063/10.0014014},
}

@misc{Besproswanny2025,
      title={Temperature dependence of surface superconductivity in t-{PtBi$_2$}}, 
      author={Julia Besproswanny and Sebastian Schimmel and Yanina Fasano and Grigory Shipunov and Saicharan Aswartham and Danny Baumann and Bernd Büchner and Christian Hess},
      year={2025},
      eprint={2507.10187},
      archivePrefix={arXiv},
      primaryClass={cond-mat.supr-con},
      url={https://arxiv.org/abs/2507.10187}, 
}

@article{Changdar2025,
      title={ Topological nodal $i$-wave superconductivity in {PtBi$_2$}}, 
      author={Changdar, S. and Suvorov, O. and Kuibarov, A. and Thirupathaiah, S. and Shipunov, G. and Aswartham, S. and Wurmehl, S. and Kovalchuk, I. and Koepernik, K. and Timm, C. and Büchner, B. and Fulga, I.C. and Borisenko, S. and van den Brink, J.},
      year={2025},
      journal={Nature},
      volume={647},
      pages = {613},
      doi={10.1038/s41586-025-09712-6}
}

@article{Feng2019,
  title = {Rashba-like spin splitting along three momentum directions in trigonal layered {PtBi$_2$}},
  author = {Feng, Ya and Jiang, Qi and Feng, Baojie and Yang, Meng and Xu, Tao and Liu, Wenjing and Yang, Xiufu and Arita, Masashi and Schwier, Eike F. and Shimada, Kenya and Jeschke, Harald O. and Thomale, Ronny and Shi, Youguo and Wu, Xianxin and Xiao, Shaozhu and Qiao, Shan and He, Shaolong},
  journal = {Nature Communications},
  volume = {10},
  pages = {4765},
  year = {2019},
  doi = {10.1038/s41467-019-12805-2},
}

@book{Gopal1966,
  title={Specific Heats at Low Temperatures},
  author={Gopal, E.S.R.},
  isbn={9780608057583},
  lccn={65011339},
  series={International cryogenics monograph series},
  url={https://books.google.de/books?id=jAhRAAAAMAAJ},
  year={1966},
  publisher={Plenum Press},
  doi={https://doi.org/10.1007/978-1-4684-9081-7}
}

@article{Gao2018,
  title = {A possible candidate for triply degenerate point fermions in trigonal layered {PtBi$_2$}},
  author = {Gao, W. and Zhu, X. and Zheng, F. and Wu, M. and Zhang, J. and Xi, C. and Zhang, P. and Zhang, Y. and Hao, N. and Ning, W. and Tian,  M. },
  journal = {Nat. Commun.},
  volume = {9},
  pages = {3249},
  year = {2018},
  doi = {https://doi.org/10.1038/s41467-018-05730-3}
}

@article{Klein2010,
   author = {Klein, T. and Braithwaite, D. and Demuer, A. and Knafo, W. and Lapertot, G. and Marcenat, C. and Rodi\`ere, P. and Sheikin, I. and Strobel, P. and Sulpice, A. and Toulemonde, P.},
   title = {Thermodynamic phase diagram of {$\text{Fe}({\text{Se}}_{0.5}{\text{Te}}_{0.5})$} single crystals in fields up to 28 tesla},
   journal = {Phys. Rev. B},
   volume = {82},
   pages = {184506},
   year = {2010},
   doi = {10.1103/PhysRevB.82.184506},
}

@article{Kudo2021,
   author = {Kudo, Kazutaka and Nguyen, Hoang Yen and Oh, Chang-geun and Takaki, Kensuke and Nohara, Minoru},
   title = {Superconductivity of the Stuffed {CdI$_2$}-type {Pt$_{1+x}$Bi$_2$}},
   journal = {J. Phys. Soc. Japan},
   volume = {90},
   number = {6},
   pages = {063706},
   year = {2021},
   doi = {10.7566/JPSJ.90.063706},
}

@article{Kuibarov2024,
  title = {Evidence of superconducting {Fermi} arcs},
  author = {Kuibarov,A. and Suvorov,O. and Vocaturo,R. and Fedorov,A. and Lou,R. and Merkwitz, L. and Voroshnin,V. and Facio,J.I. and Koepernik,K. and Yaresko,A. and Shipunov,G. and Aswartham,S. and van den Brink,J. and B\"uchner,B.  and Borisenko,S. },
  journal = {Nature},
  volume = {626},
  pages = {294},
  year = {2024},
  doi = {https://doi.org/10.1038/s41586-023-06977-7},
}

@article{Kvitnitskaya2025,
doi = {10.1088/1361-6668/ae61fb},
url = {https://doi.org/10.1088/1361-6668/ae61fb},
year = {2026},
month = {may},
publisher = {IOP Publishing},
volume = {39},
number = {5},
pages = {055004},
author = {Kvitnitskaya, O E and Harnagea, L and Shipunov, G and Aswartham, S and Kovalchuk, I and Fisun, V V and Efremov, D V and Büchner, B and Naidyuk, Yu G},
title = {Point-contact enhanced superconductivity in trigonal {PtBi$_2$}: quest for the origin of {‘high-T$_c$’}},
journal = {Superconductor Science and Technology}
}

@article{Okamoto1991,
    author = {H. Okamoto},
    title = {The {Bi-Pt} (bismuth-platinum) system},
    journal = {J. Phase Equilib.},
    volume = 12,
    pages = 207,
    year = 1991,
    doi = {https://doi.org/10.1007/BF02645718}}

@article{Schimmel2023,
  title = {Surface superconductivity in the topological {Weyl} semimetal t-{PtBi$_2$}},
  author = {Schimmel, S. and Fasano, Y. and Hoffmann, S. and Besproswanny, J. and Corredor Bohorquez, L. T. and Puig, J. and Elshalem, B.-C. and Kalisky, B. and Shipunov, G. and Baumann, D. and Aswartham, S. and Büchner, B. and Hess, C.},
  journal = {Nat. Commun.},
  volume = {15},
  pages = {9895},
  year = {2024},
  url = {https://doi.org/10.1038/s41467-024-54389-6},
}

@article{Shipunov2020,
  title = {Polymorphic {PtBi$_{2}$}: {Growth}, structure, and superconducting properties},
  author = {Shipunov, G. and Kovalchuk, I. and Piening, B. R. and Labracherie, V. and Veyrat, A. and Wolf, D. and Lubk, A. and Subakti, S. and Giraud, R. and Dufouleur, J. and Shokri, S. and Caglieris, F. and Hess, C. and Efremov, D. V. and B\"uchner, B. and Aswartham, S.},
  journal = {Phys. Rev. Mater.},
  volume = {4},
  pages = {124202},
  year = {2020},
  doi = {10.1103/PhysRevMaterials.4.124202}}

@article{Takaki2022,
author = {Takaki, K. and Yamamoto, M. and Nakajima, M. and Takeuchi,T. and Nguyen,H.Y. and Nohara, M. and Kishioji, Y. and Fujii, T. and Yoshino, K. and Miyasaka, S. and Kudo, K.},
title = "{Enhanced Superconductivity in Close Proximity to Polar-Nonpolar Structural Phase Transition in Se/Te-Substituted PtBi$_2$}",
journal = {Journal of the Physical Society of Japan},
volume = {91},
number = {3},
pages = {034703},
year = {2022},
doi = {10.7566/JPSJ.91.034703}}

@article{Thirupathaiah2018,
  title = {Possible origin of linear magnetoresistance: Observation of {Dirac} surface states in layered {PtBi$_2$}},
  author = {Thirupathaiah,S. and Kushnirenko,Y. and Haubold,E. and Fedorov,A. V. and Rienks, E. D. L. and Kim, T. K. and Yaresko,A. N. and Blum,C. G. F. and Aswartham,S. and B\"uchner,B. and Borisenko,S. V.},
  journal = {Phys. Rev. B},
  volume = {97},
  pages = {035133},
  year = {2018},
  doi = {https://doi.org/10.1103/PhysRevB.97.035133}}

@book{TinkhamSC,
  title={Introduction to {Superconductivity}},
  author={Tinkham, M.},
  isbn={9780486435039},
  series={Dover Books on Physics Series},
  year={1996},
  publisher={Dover Publications}
}

@article{Veyrat2022,
  title = {Superconductivity and topology in trigonal-{PtBi$_2$}},
  author = {Veyrat, A.},
  journal = {Ph.D. thesis, Technische Universit\"at Dresden},
  volume = {},
  pages = {},
  year = {2022},
  url = {https://nbn-resolving.org/urn:nbn:de:bsz:14-qucosa2-792057},
}

@article{Veyrat2023,
  title = {{Berezinskii-Kosterlitz-Thouless} transition in the type {I} {Weyl} semimetal {PtBi$_2$}},
  author = {Veyrat, A. and Labracherie, V. and Bashlakov, D.L. and Caglieris, F. and Facio, J.I. and Shipunov, G. and Charvin, T. and Acharya, R. and Naidyuk, Y. and Giraud, R. and van den Brink, J. and B\"uchner, B. and Hess, C. and Aswartham, S. and Dufouleur, J. },
  journal = {Nano Lett.},
  volume = {23},
  pages = {1229},
  year = {2023},
  doi = {https://doi.org/10.1021/acs.nanolett.2c04297},
}

@article{Vocaturo2024,
  title = {Electronic structure of the surface-superconducting {Weyl} semimetal {${\mathrm{PtBi}}_{2}$}},
  author = {Vocaturo, Riccardo and Koepernik, Klaus and Facio, Jorge I. and Timm, Carsten and Fulga, Ion Cosma and Janson, Oleg and van den Brink, Jeroen},
  journal = {Phys. Rev. B},
  volume = {110},
  issue = {5},
  pages = {054504},
  numpages = {12},
  year = {2024},
  month = {Aug},
  publisher = {American Physical Society},
  doi = {10.1103/PhysRevB.110.054504},
  url = {https://link.aps.org/doi/10.1103/PhysRevB.110.054504}
}

@article{Werthamer1966,
  title = "{Temperature and Purity Dependence of the Superconducting Critical Field, ${H}_{c2}$. III. Electron Spin and Spin-Orbit Effects}",
  author = {Werthamer, N. R. and Helfand, E. and Hohenberg, P. C.},
  journal = {Phys. Rev.},
  volume = {147},
  issue = {1},
  pages = {295--302},
  numpages = {0},
  year = {1966},
  month = {Jul},
  publisher = {American Physical Society},
  doi = {10.1103/PhysRev.147.295},
  url = {https://link.aps.org/doi/10.1103/PhysRev.147.295}
}

@article{Wu2020,
  title = {Huge linear magnetoresistance due to open orbits in $\gamma$-{PtBi$_2$}},
  author = {Wu, B. and Barrena, V. and Suderow, H. and Guillam\'on, I.},
  journal = {Phys. Rev. Res.},
  volume = {2},
  pages = {022042(R)},
  year = {2020},
  doi = {https://doi.org/10.1103/PhysRevResearch.2.022042},
}

@article{Xing2020,
  title = {Topological behavior and {Zeeman} splitting in trigonal {PtBi$_{2-x}$} single crystals},
  author = {Xing, L. and Chapai, R. and Nepal, R. and Jin, R. },
  journal = {npj Quantum Materials},
  volume = {5},
  pages = {10},
  year = {2020},
  doi = {10.1038/s41535-020-0213-9}}

@article{Xu2016,
  title = {Synthesis, physical properties, and band structure of the layered bismuthide {PtBi$_2$}},
  author = {Xu, C.Q. and Xing, X.Z. and Xu, X. and Li, B. and Chen, B. and Che, L.Q. and  Lu, X. and Dai, J. and Shi, Z.X.},
  journal = {Phys. Rev. B},
  volume = {94},
  pages = {165119},
  year = {2016},
  doi = {https://doi.org/10.1103/PhysRevB.94.165119}
}

@article{Yuan2021,
doi = {10.1088/1361-6463/abe78b},
year = {2021},
volume = {54},
pages = {215302},
author = {Yuan, Yahua and Duan, Yuxia and Wang, Zhongwang and Sun, Jian},
title = {Filamentary superconductivity in wrinkled {PtSe$_2$}},
journal = {J. Phys. D: Appl. Phys.}
}

@article{Zabala2024,
   title={Enhanced weak superconductivity in trigonal $\gamma$-{PtBi$_2$}},
   author={Zabala, J and Correa, V F and Castro, F J and Pedrazzini, P},
   journal={Journal of Physics: Condensed Matter},
   volume={36},
   pages={285701}, 
   year={2024},
   DOI={10.1088/1361-648x/ad3878},
}

@manual{CrysAlisPRO,
  title        = "{CrysAlis PRO}",
  organization = "{Agilent Technologies Ltd, Yarnton, Oxfordshire, England}",
  year         = 2014,
}

@article{Jana2006,
  title = "{Crystallographic Computing System JANA2006: General features}",
  author = {Václav Petříček and Michal Dušek and Lukáš Palatinus},
  journal = {Zeitschrift für Kristallographie - Crystalline Materials},
  volume = {229},
  pages = {345--352},
  number = {5},
  year = {2014},
  doi = {doi:10.1515/zkri-2014-1737},
  url = {https://doi.org/10.1515/zkri-2014-1737},
}

@article{superflip,
  title={{SUPERFLIP} - a computer program for the solution of crystal structures by charge flipping in arbitrary dimensions},
  author={Palatinus, Lukas and Chapuis, Gervais},
  journal={Journal of Applied Crystallography},
  volume={40},
  number={4},
  pages={786--790},
  year={2007},
  url = {https://journals.iucr.org/j/issues/2007/04/00/db5026/db5026.pdf}
}

@article{Naidyuk2018,
  title={Anatomy of point-contact {Andreev} reflection spectroscopy from the experimental point of view},
  author={Naidyuk, Yu.G. and Gloos, K. },
  journal={Fiz. Nizk. Temp},
  volume={44},
  pages={343},
  year={2018},
  doi = { https://doi.org/10.1063/1.5030447}
}

@misc{huang2025,
      title={Sizable superconducting gap and anisotropic chiral topological superconductivity in the {Weyl} semimetal {PtBi$_2$}}, 
      author={Xiaochun Huang and Lingxiao Zhao and Sebastian Schimmel and Julia Besproswanny and Patrick Härtl and Christian Hess and Bernd Büchner and Matthias Bode},
      year={2025},
      eprint={2507.13843},
      archivePrefix={arXiv},
      primaryClass={cond-mat.supr-con},
      url={https://arxiv.org/abs/2507.13843}, 
}

@article{Palumbo2025,
        title={Interplay between inversion and translation symmetries in trigonal {PtBi$_2$}}, 
        author={Palumbo, S. and Cornaglia, P.S. and Facio, J.I.},
        year={2025},
        journal = {Phys. Rev. B},
        volume = {112},
        pages = {205125},
        publisher = {American Physical Society},
        doi = {10.1103/bb1q-4qt5}
}

@article{Samukawa2025,
  title = {Observation of two-level critical-state in the {van-der-Waals} superconductor {$\mathrm{Pt}{({\mathrm{Bi}}_{1\ensuremath{-}x}{\mathrm{Se}}_{x})}_{2}$}},
  author = {Samukawa, Y. and Maeda, M. and Jiang, N. and Nakamura, R. and Watanabe, M. and Takaki, K. and Moriyasu, Y. and Ikushima, T. and Kida, T. and Hagiwara, M. and Nakamura, T. and Okada, Y. and Kudo, K. and Niimi, Y.},
  journal = {Phys. Rev. B},
  volume = {111},
  issue = {6},
  pages = {064506},
  numpages = {9},
  year = {2025},
  month = {Feb},
  publisher = {American Physical Society},
  doi = {10.1103/PhysRevB.111.064506},
  url = {https://link.aps.org/doi/10.1103/PhysRevB.111.064506}
}

@article{Naidyuk2020,
    author = {Naidyuk, Yu.G. and Kvitnitskaya, O. E. and Efremov, D. V. and Drechsler, S.-L.},
    title = {Enhanced critical temperatures in the strongly overdoped iron-based superconductors {AFe$_2$As$_2$ (A=K, Cs, Rb)} observed by point contacts},
    journal = {Low Temperature Physics},
    volume = {46},
    number = {11},
    pages = {1070-1073},
    year = {2020},
    month = {11},
    issn = {1063-777X},
    doi = {10.1063/10.0002149},
    url = {https://doi.org/10.1063/10.0002149},
}

@article{Naidyuk2016,
  title = {Doubling of the critical temperature of {FeSe} observed in point contacts},
  author = {Naidyuk, Yu.G. and Fuchs, G. and Chareev, D. A. and Vasiliev, A. N.},
  journal = {Phys. Rev. B},
  volume = {93},
  issue = {14},
  pages = {144515},
  numpages = {3},
  year = {2016},
  month = {Apr},
  publisher = {American Physical Society},
  doi = {10.1103/PhysRevB.93.144515},
  url = {https://link.aps.org/doi/10.1103/PhysRevB.93.144515},
}

@ARTICLE{fesuma,
  title     = {Fermi surface tomography},
    author    = {Borisenko, Sergey and Fedorov, Alexander and Kuibarov,                     Andrii
               and Bianchi, Marco and Bezguba, Volodymyr and Majchrzak, Paulina
               and Hofmann, Philip and Baumg{\"a}rtel, Peter and Voroshnin,
               Vladimir and Kushnirenko, Yevhen and S{\'a}nchez-Barriga, Jaime
               and Varykhalov, Andrei and Ovsyannikov, Ruslan and Morozov, Igor
               and Aswartham, Saicharan and Feia, Oleh and Harnagea, Luminita
               and Wurmehl, Sabine and Kordyuk, Alexander and Yaresko,
               Alexander and Berger, Helmuth and B{\"u}chner, Bernd},
  journal   = {Nat. Commun.},
  publisher = {Springer Science and Business Media LLC},
  volume    =  {13},
  number    =  {1},
  pages     = {4132},
  month     =  {jul},
  year      =  {2022},
  copyright = {https://creativecommons.org/licenses/by/4.0},
  doi = {https://doi.org/10.1038/s41467-022-31841-z}
}

@misc{supp,
  note = {See Supplemental Material at
    URL-will-be-inserted-by-publisher.}
}

@article{Kuibarov2025,
  title = {Measuring superconducting arcs by angle-resolved photoemission spectroscopy},
  author = {Kuibarov, A. and Changdar, S. and Fedorov, A. and Lou, R. and Suvorov, O. and Misheneva, V. and Harnagea, L. and Kovalchuk, I. and Wurmehl, S. and B\"uchner, B. and Borisenko, S.},
  journal = {Phys. Rev. B},
  volume = {112},
  issue = {14},
  pages = {144518},
  numpages = {7},
  year = {2025},
  month = {Oct},
  publisher = {American Physical Society},
  doi = {10.1103/5q36-wgl9},
  url = {https://link.aps.org/doi/10.1103/5q36-wgl9}
}
\end{document}